\documentclass[12pt,preprint]{aastex}
\begin{document}

\title{Inner crusts of neutron stars in strongly quantising magnetic fields}
\author{Rana Nandi\altaffilmark{1} and 
Debades Bandyopadhyay\altaffilmark{1}} 
\affil{Astroparticle Physics and Cosmology Division, Saha Institute of Nuclear 
Physics, 1/AF Bidhannagar, Kolkata-700064, India}
\altaffiltext{1}{Centre for Astroparticle Physics, Saha Institute of Nuclear
Physics, 1/AF Bidhannagar, Kolkata-700064, India}
\author{Igor N. Mishustin and Walter Greiner} 
\affil{Frankfurt Institute for Advanced Studies (FIAS), J. W. Goethe 
Universit\"at, Ruth Moufang Strasse 1, 60438 Frankfurt am Main, Germany}

\begin{abstract}
We study the properties and stability of nuclei in the inner crust of neutron 
stars in the
presence of strong magnetic fields $\sim 10^{17}$ G. Nuclei coexist with a 
neutron gas and reside in a uniform gas of electrons in the inner crust. This 
problem is investigated within the Thomas-Fermi model. We extract the 
properties of nuclei based on the subtraction procedure of Bonche, Levit and 
Vautherin. The phase space modification of electrons due to Landau quantisation
in the presence of strong magnetic fields leads to the 
enhancement of electron as
well as proton fractions at lower densities $\sim 0.001$ fm$^{-3}$. We find the 
equilibrium nucleus at each average baryon density by minimising the free 
energy and show that ,in the presence of strong magnetic 
fields, it is lower than that in the field free case. 
The size of the spherical cell that
encloses a nucleus along with the neutron and electron gases becomes smaller
in strong magnetic fields compared with the zero field case. Nuclei with larger
mass and atomic numbers are obtained in the presence of strong magnetic fields
as compared with cases of zero field.
\end{abstract}


\section{Introduction}
Strong surface magnetic fields $\sim 10^{12}$ G are found to exist in pulsars. 
Even stronger surface magnetic fields $\geq 10^{15}$ G were predicted by 
observations on soft gamma ray repeaters and anomalous x-ray pulsars 
\citep{kouvel98,kouvel99}. The latter class of neutron stars with very intense 
magnetic fields is known as magnetars 
\citep{thomp93,thomp96}. On the other hand, the interior magnetic field could 
be much higher than the surface 
field. The limiting interior field might be estimated from the scalar virial
theorem \citep{Shapiro}.  For a typical neutron star mass 1.5$M_{\odot}$ and
radius 15 km, the interior field could be as high as $\sim 10^{18}$ G.  

Such strong magnetic fields quantise the motion of charged particles 
perpendicular to the field \citep{Land}. The effects of the phase space 
modification due to the Landau quantisation were studied on the composition and
equation of state (EoS) in neutron stars extensively. Lai 
and
Shapiro extended the Baym, Pethick and Sutherland (BPS) model \citep{bps} to 
the magnetic field case and obtained equilibrium nuclei and the EoS in the
outer crust in the presence of strong magnetic fields \citep{Shapiro}. The 
composition and EoS in the core of neutron stars in the presence of strongly 
quantising magnetic fields
were investigated within a relativistic field theoretical model by Chakrabarty
et al. \citep{cbs,bcp}. The transport properties such as thermal and electrical
conductivities of neutron star crusts in magnetic fields were studied by 
several groups \citep{Yak,Hern}. Recently the magnetised neutron star crust was
studied using the Thomas-Fermi model and Baym-Bethe-Pethick \citep{bbp}
and Harrison-Wheeler EoS for nuclear matter \citep{Nag}.   

In the outer crust of a neutron star, neutrons and protons are bound inside
nuclei and immersed in a uniform background of relativistic electron gas. As
the density increases, nuclei become more and more neutron rich. Neutrons start
to drip out of nuclei at a density $\sim 4 \times 10^{11}$ g/cm$^3$. This is 
the beginning of the inner crust. The matter in the inner crust is made of 
nuclei embedded in a neutron gas along with the uniform electron gas. Further
the matter is in $\beta$-equilibrium and maintains charge neutrality. Nuclei
are also in mechanical equilibrium with the neutron gas. The 
properties of nuclei in the inner crusts of neutron stars in zero magnetic 
field were studied by different groups. The early studies of the inner crust
matter were based on the extrapolations of the semiempirical mass formula to 
the free neutron gas regime \citep{Lan,Sato}. Baym, Bethe and Pethick 
considered the reduction of the nuclear surface energy due to the free neutron 
gas in their calculation \citep{bbp}. The study of nuclei in the neutron star 
crust was carried out using the energy density of a many body system by 
Negele and Vautherin \citep{Neg}. With increasing density in the inner crust, 
unusual nuclear shapes
might appear there \citep{Rav,Oya}. The properties of nuclei in the
inner crust were also investigated using a relativistic field theoretical model
\citep{Che}.  

There are two important aspects of the problem when nuclei are immersed in 
a neutron gas. On the one hand we have to deal with the coexistence of two 
phases of nuclear matter - denser phase inside a nucleus and low density phase 
outside it, in a thermodynamical consistent manner. On the other hand, the 
determination of the surface energy of the interface between two phases with
good accuracy is needed. It was shown that this problem could be solved using
the subtraction procedure of Bonche, Levit and Vautherin \citep{Bon1,Bon2,Sur}.
The properties of a nucleus are isolated from nucleus plus neutron gas in a 
temperature dependent Hartree-Fock theory
using the subtraction procedure. This
same method was extended to isolated nuclei embedded in a neutron gas \citep{De}
as well as nuclei in the inner crust at zero temperature \citep{Sil}. This
shows that it would
be worth studying the properties of nuclei in the inner crust in the presence 
of strongly quantizing magnetic field relevant to magnetars using the 
subtraction procedure.

Recently the stability of nuclei embedded in an electron gas was investigated
within a relativistic mean field model in zero magnetic field \citep{Mis}. It 
was observed in their calculation that nuclei became more stable against 
$\alpha$
decay and spontaneous fission with increasing electron number density. It is 
worth mentioning here that the electron number density is enhanced in the 
presence of strong magnetic fields due to Landau quantisation compared with the
zero field case. The question is whether the nuclear system in the inner crust 
of magnetars
would be more stable than those of the field free case. This is the focus of 
our calculation in this article.     

The paper is organised in the following way. In section 2, the formalism for
the calculation of nuclei of the inner crust immersed in a neutron as well as 
an electron gas in the presence of strongly quantising magnetic fields 
is described. Results of our calculation are discussed in section 3. Section 4 
contains the summary and conclusions. 

\section{Formalism}
We investigate the properties of nuclei and their stability in the inner crust 
in 
the presence of strong magnetic fields using the Thomas-Fermi (TF) model. In 
this case nuclei are
immersed in a nucleonic gas as well as a uniform background of electrons 
and may be arranged in a lattice. Each lattice volume is
replaced by a spherical cell with a nucleus at its center in the Wigner-Seitz 
(WS) approximation.  Each cell is taken to be charge neutral such that
the number of electrons is equal to the number of protons in it. The Coulomb 
interaction between cells is neglected. Electrons are assumed to be uniformly 
distributed within a cell. The system maintains the $\beta$-equilibrium. 
We assume that the system is placed in a uniform magnetic field. Electrons are 
affected by strongly quantizing magnetic fields. Protons in the cell are 
affected by magnetic fields only through the charge neutrality condition. The 
interaction of nuclear magnetic moment with the field is negligible in a 
magnetic field $\sim 10^{17}$ G \citep{Lat}. 

The calculation below is performed in a zero temperature TF model. In the WS 
cell, a nucleus is located at the centre and immersed in a low density neutron 
gas whereas 
protons are trapped in the nucleus. However, the spherical cell does not 
define a nucleus. The nucleus is realised after subtraction of the gas part 
from the cell as shown by Bonche, Levit and Vautherin \citep{Bon1,Bon2}. In an
earlier calculation, it was demonstrated that the TF formalism at finite 
temperature gave two solutions \citep{Sur}. One solution corresponds to 
the nucleus plus neutron gas and the second one represents only the 
neutron gas. The density profiles of the nucleus plus neutron gas as well
as that of the neutron gas are obtained self-consistently in the TF 
formalism. Finally the nucleus is obtained as the difference of two 
solutions. This formalism is adopted in our calculation at zero temperature as
described below. 

The nucleus plus gas solution coincides with the gas solution at large
distance leading to the definition of the thermodynamic potential 
($\Omega_N$) of the nucleus as \citep{Bon1,Bon2}
\begin{equation}
\Omega_N = \Omega_{NG} - \Omega_G~,
\end{equation} 
where $\Omega_{NG}$ is the thermodynamic potential of the nucleus plus gas 
phase and $\Omega_G$ is that of the gas only.
The thermodynamic potential is defined as
\begin{equation}
\Omega = {\cal{F}} - \sum_{q=n,p} \mu_q A_q~,
\end{equation} 
where $\mu_q$ and $A_q$ are the chemical potential and number of q-th 
species, respectively.
The free energy is given by
\begin{equation}
{\cal{F}}(n_b,Y_p) = \int [{\cal{H}} + \varepsilon_c + \varepsilon_e] d{\bf r}~,
\end{equation}  
where $\cal{H}$ is nuclear energy density functional, $\varepsilon_c$ is the 
Coulomb 
energy density and $\varepsilon_e$ is the energy density of electrons. The free
energy is a function of average baryon density ($n_b$) and proton fraction 
($Y_p$).
The nuclear energy density is calculated using the SKM$^*$ nucleon-nucleon 
interaction and given by \citep{Sil,Bra,Stone} 
  \begin{eqnarray}
 {\cal H}(r)&=&\frac{\hbar^2}{2m_n^*}\tau_n 
 +\frac{\hbar^2}{2m_p^*}\tau_p+
 \frac{1}{2}t_0\left[\left(1+\frac{x_0}{2}\right)\rho^2-\left(x_0+
 \frac{1}{2}\right)\left(\rho^2_n+\rho^2_p\right)\right]\nonumber\\
 &&-\frac{1}{16}\left[t_2\left(1+\frac{x_2}{2}\right)
 -3t_1\left(1+\frac{x_1}{2}\right)\right](\nabla \rho)^2\nonumber\\
 &&-\frac{1}{16}\left[3t_1\left(x_1+\frac{1}{2}\right)
 +t_2\left(x_2+\frac{1}{2}
 \right)\right]\left[(\nabla\rho_n)^2+(\nabla\rho_p)^2\right]\nonumber\\
 &&+\frac{1}{12}t_3\rho^\alpha\left[\left(1+\frac{x_3}{2}\right)\rho^2
 -\left(x_3+\frac{1}{2}\right)\left(\rho_n^2+\rho_p^2\right)\right]~.
 \end{eqnarray} 
The first two terms of the nuclear energy density are kinetic energy densities
of neutrons and protons, respectively. The third term originates from the zero
range part of the Skyrme interaction whereas the last term is the contribution
of the density dependent part of the nucleon-nucleon interaction. The effective mass of nucleons is given by 
 \begin{eqnarray}
 \frac{m}{m_q^*(r)}&=&1+\frac{m}{2\hbar^2}\left\{\left[t_1\left(1+
 \frac{x_1}{2}\right)
 +t_2\left(1+\frac{x_2}{2}\right)\right]\right.\rho \nonumber\\
 && +\left.\left[t_2\left(x_2+\frac{1}{2}\right)
 -t_1\left(x_1+\frac{1}{2}
 \right)\right] \rho_q\right\}~,
 \end{eqnarray}
where total baryon density is $\rho = \rho_n + \rho_p$.

The Coulomb energy densities for the NG and G phases are:
 \begin{eqnarray}
\varepsilon^{NG}_c (r) &=&\frac{1}{2} (n_{NG}^p(r)-n_e) 
\int \frac{e^2}{\mid{\bf r}-{\bf r^{\prime}}\mid} 
(n_{NG}^p(r')-n_e)d{\bf r'}\nonumber\\
\varepsilon^{G}_c (r) &=&\frac{1}{2} (n_{G}^p(r)-n_e) 
\int \frac{e^2}{\mid{\bf r}-{\bf r'}\mid} 
(n_{G}^p(r')-n_e)d{\bf r'}\nonumber\\
&& +n_{N}^p(r)
\int \frac{e^2}{\mid{\bf r}-{\bf r^{\prime}}\mid} 
(n_{G}^p(r')-n_e)d{\bf r'}~,
\end{eqnarray}
where $n_{NG}^p$ and $n_{G}^p$ are proton densities in the nucleus plus gas and
only gas. Here the coulomb energy densities ($\varepsilon_c$) represent the 
direct part. We do not consider the exchange part because its contribution is
small. 

So far the formalism described above is applicable for the zero magnetic field.
However, we study the effects of magnetic fields on electrons which , in turn, 
influence the properties of nuclei in the inner crust. The 
Coulomb energy 
density and the energy density of electrons in Eq. (3) would be modified in 
strongly quantising magnetic fields. In the 
presence of a magnetic field, the motion of electrons is quantized in the 
plane perpendicular to the field. Protons in nuclei would be influenced by a 
magnetic field through the charge neutrality condition. 
We take the magnetic field ($\overrightarrow{B}$) along 
Z-direction and assume that it is uniform throughout the inner crust.
If the field strength exceeds a critical value 
$B_c=m_e^2/e\simeq 4.414\times 10^{13}$G, then electrons become relativistic
\citep{Lai}. 
The energy eigenvalue of relativistic electrons in quantizing magnetic field is
given by
\begin{equation}
 E_e(\nu,p_z)=\left[p_z^2+m_e^2(1+2\nu B_{*})\right]^{1/2} ~,
\end{equation}
where $p_z$ is the Z-component of momentum, $B_{*}=B/B_c$, $\nu$ is 
the Landau quantum number. The Fermi momentum of electrons, $p_{f_{e}}$, is
obtained from 
\begin{equation}
\mu_e = \left[p_{f_e}(\nu)^2+m_e^2(1+2\nu B_{*})\right]^{1/2} - V^c(r)~,
\end{equation}
where $V^c(r)$ is the direct part of the single particle Coulomb potential.

The number density of electrons in a magnetic field is calculated as
\begin{equation}
 n_e=\frac{eB}{2\pi^2}\sum_{\nu=0}^{\nu_{max}}g_{\nu}p_{f_{e}}(\nu)~.
\end{equation}
Here the spin degeneracy is 
$g_\nu = 1$ for the lowest Landau level ($\nu=0$) and $g_\nu = 2$ for all
other levels.

The maximum Landau quantum number ($\nu_{max}$) is given by
\begin{equation}
 \nu_{max}=\frac{\left(\mu_e + V^c(r)\right)^2-m_e^2}{2eB}~.
\end{equation}
The energy density of electrons is obtained from,
\begin{eqnarray}
\varepsilon_e=\frac{eB}{2\pi^2}\sum_0^{\nu_{max}}g_\nu \int_0^{p_{f_{e}}(\nu)}
E_e(\nu,p_z)dp_z~.
\end{eqnarray}

We minimise the thermodynamic potential in the TF approximation with the 
condition of number conservation of each species. The density profiles of 
neutrons and protons with or without magnetic fields are obtained from
\begin{eqnarray}
\frac{\delta \Omega_{NG}}{\delta n_{NG}^q}&=&0~,\nonumber\\ 
\frac{\delta \Omega_{G}}{\delta n_{G}^q}&=&0~.
\end{eqnarray}
This results in the following coupled equations \citep{Sil,De}
  \begin{eqnarray}
(3{\pi}^2)^{2 \over 3} \frac{\hbar^2}{2m_q^*}({n_{NG}^q})^{2 \over 3} + V_{NG}^q
+V_{NG}^c(n_{NG}^p,n_e)&=&\mu_q~,\nonumber\\
(3{\pi}^2)^{2 \over 3} \frac{\hbar^2}{2m_q^*} ({n_{G}^q})^{2 \over 3} + V_{G}^q
+V_{G}^c(n_e)&=&\mu_q~,
\end{eqnarray}
where $m_q^*$ is the effective mass of q-th species , $V_{NG}^q$ and $V_G^q$ 
are the single particle potentials of nucleons in the nucleus plus gas as well
as gas phases \citep{Bra}. On the other hand, $V_{NG}^c$ and $V_G^c$ are
direct parts of the single particle Coulomb potential corresponding to
the nucleus plus gas and only gas solutions and both are given by 
\begin{equation}
  V^c(r) = \int\left[n_{NG}^p(r')-n_e\right]\frac{e^2}{\mid{{\bf r}
-{\bf r'}}\mid} d{\bf r'}~.
\end{equation}

The average chemical potential for q-th nucleon is
\begin{equation}
\mu_q=\frac{1}{A_q}\int [(3{\pi}^2)^{2 \over 3} \frac{\hbar^2}{2m^*_q}
{(n^q_{NG})^{2 \over 3}} + V_{NG}^q(r)+V_{NG}^c(r)]\rho_{NG}^q(r) d{\bf r}
\end{equation}
where $A_q$ refers to $N_{cell}$ or $Z_{cell}$ of the cell which is defined by 
the average
baryon density $n_b$ and proton fraction $Y_p$. The $\beta$-equilibrium 
condition is written as
\begin{equation}
\mu_n = \mu_p + \mu_e~.
\end{equation}
The average electron chemical potential in magnetic fields is given by
\begin{equation}
\mu_e = \left[p_{f_e}(\nu)^2+m_e^2(1+2\nu B_{*})\right]^{1/2} - <V^c(r)>~,
\end{equation}
where $<V^c(r)>$ denotes the average of the single particle Coulomb 
potential. 

Density profiles of neutrons and protons in the cell are constrained as
\begin{eqnarray}
Z_{cell}&=&\int \rho_p^{NG}(r)d{\bf r}~,\nonumber\\
N_{cell}&=&\int \rho_n^{NG}(r)d{\bf r}~,
\end{eqnarray}
where
$N_{cell}$ and $Z_{cell}$ are neutron and proton numbers in the cell,
respectively.

Finally, number of neutrons ($N$) and protons ($Z$) in a nucleus with mass 
number $A = N + Z$ are obtained using the subtraction procedure as
\begin{eqnarray}
Z&=&\int \left[\rho_p^{NG}(r)-\rho_p^G(r)\right]d{\bf r}~,\nonumber\\
N&=&\int \left[\rho_n^{NG}(r)-\rho_n^G(r)\right]d{\bf r}~.
\end{eqnarray}

\section{Results and Discussion}       
We find out the equilibrium nucleus at each density point minimising the free
energy of the system within a WS cell maintaining charge neutrality and
$\beta$-equilibrium. The variables of this problem are the average baryon 
density ($n_b$), the proton fraction ($Y_p$) and the radius of a cell ($R_c$).
For a fixed value of $n_b$, $Y_p$ and $R_c$, the total number of nucleons 
($A_{cell}$) is given by $A_{cell} = V_{cell}n_b$
where the volume of a cell is $V_{cell} = 4/3 \pi R_C^3$. The proton number in
the cell is $Z_{cell} = Y_p n_b V_{cell}$ and the neutron number is 
$N_{cell} = A_{cell} - Z_{cell}$. Now we obtain density profiles of neutrons and
protons in the cell using Eqs. (13) and (18) at a given average baryon density
and $Y_p$.
Consequently, we calculate chemical 
potentials of neutrons and protons and free energy per nucleon.
Next we vary the proton fraction, calculate chemical potentials and density 
profiles and obtain the $\beta$-equilibrium in the cell.
Finally we adust the cell size ($R_C$) and repeat the above mentioned steps to 
get the minimum of the free energy. These values of $Y_p$ and $R_C$ are then 
used to calculate neutron and proton numbers in a nucleus at an average baryon 
density corresponding to the free energy minimum with the help of Eq. (19). 
This procedure is repeated for each average baryon density. 

The minimum of the free energy originates from the interplay of different
contributions. The free energy per nucleon is given by
\begin{equation}
F/A = e_N + e_{lat} + e_{ele}~.
\end{equation}
The nuclear energy including the Coulomb interaction among protons is denoted
by $e_N$, $e_{lat}$ is the lattice energy which involves the Coulomb 
interaction between electrons and protons and the electron kinetic energy is
$e_{ele}$. The free energy per nucleon in the presence of magnetic field 
$B = 4.414 \times 10^{16}$ G is shown as a function of the cell size
for an average baryon density $n_b =0.008$ fm $^{-3}$ in Fig. 1. 
We note that the nuclear energy increases with $R_C$. On the other
hand, the lattice energy and electron kinetic energy both decrease with 
increasing cell size. The competition of $e_N$ with the sum of $e_{lat}$ and
$e_{ele}$ determines the free energy minimum. The cell radius corresponding to 
the free energy minimum is 32.1 fm for the zero field case (not shown in the
figure) and 31.9 fm for $B = 4.414 \times 10^{16}$ G. The corresponding proton 
fraction for $B = 4.414 \times 10^{16}$ G is 0.03.  

In Figure 2, the cell size corresponding to the 
free energy minimum is plotted as a function of average baryon 
density for magnetic fields $B=0$, $4.414 \times 10^{16}$, $10^{17}$
and $4.414\times 10^{17}$ G. For magnetic fields $B < 10^{17}$, several 
Landau levels are populated by electrons. Consequently we do not find any 
change in the cell size in the magnetic fields compared with the zero field 
case. However we find some change in the cell size for $B = 10^{17}$ G when 
only the zeroth Landau level is populated by electrons for $n_b \leq 0.004$, 
$fm^{-3}$ whereas first two levels are populated in the density range 0.005 to 
0.015 fm$^{-3}$. However the cell size is increased due to the
population of only zeroth Landau level in the presence of
the magnetic field $B=4.414\times 10^{17}$G compared with the zero field case. 
The size of the cell always decreases with increasing average baryon density.

The proton fraction in the presence of magnetic fields is shown as
a function of average baryon density in Fig. 3.
Protons in nuclei are affected by the Landau quantisation of electrons 
through the charge neutrality condition in a cell. For magnetic fields 
$B < 10^{17}$ G, the proton fraction is the same
as that of the zero field case over the whole density range considered here. We 
find some changes in the proton fraction below $n_b = 0.015$ fm$^{-3}$ when the
field is $10^{17}$ G. Though electrons populate the zeroth Landau 
level for $n_b \leq 0.004$ fm$^{-3}$, the proton fraction decreases 
below the corresponding proton fraction of the zero field case. However, for 
magnetic 
field $B = 4.414 \times 10^{17}$ G, the proton fraction is strongly enhanced 
due to the population of the zeroth Landau level for $n_b \leq 0.04$ fm$^{-3}$
compared with the zero field case.  

The density profile of neutrons in the nucleus plus gas (NG) as well as gas (G)
phases corresponding to $n_b = 0.02$ fm$^{-3}$
with and without magnetic fields are exhibited as a function of 
distance ($r$) within the cell in Fig. 4. The solid line denotes the
zero field case where as the dashed line represents the density profile with
the field $B=4.414 \times 10^{17}$ G. The horizontal lines imply the uniform 
gas phases in both cases. The proton fraction is 0.040 for the magnetic field 
case whereas it
is 0.022 for the zero field case. Though neutrons are not directly affected by 
the magnetic field, 
the neutron chemical potential is modified through the $\beta$-equilibrium due 
to Landau quantisation of electrons. Consequently, the number density of 
neutrons is altered. We find that the neutron density is higher in the gas 
phase for the zero field case than that of the situation with the magnetic 
field. We show the subtracted density profiles of neutrons with magnetic field 
$B = 4.414 \times 10^{17}$ G in Fig. 5. The neutron density profiles in the 
nucleus phase with and without magnetic field are different. 
Further we note that less number of neutrons 
drip out of a nucleus in the presence of the magnetic field than the situation
without the field. This may be attributed to the shift in the 
$\beta$-equilibrium in strong magnetic fields. We find a similar situation in 
the calculation of the outer crust in magnetic fields that the neutron drip 
point is shifted to higher densities \citep{Nandi}.    

Now we know the density profiles of neutron and protons in the nucleus plus gas
phase as well as in the nucleus at each average baryon density. We immediately
calculate the total number of neutrons and protons in the nucleus plus gas 
phase and in a nucleus using Eqs. (18) and (19). We  show total number 
($A_{cell}$) in a cell for magnetic fields $B= 4.414 \times 10^{16}$, $10^{17}$
and $4.414 \times 10^{17}$ G with average baryon density in Fig. 6. The dotted
line denotes the zero field case. In all cases, $A_{cell}$ growing with the 
density reaches a maximum and then decreases. Such a trend was observed in the
calculation of Negele and Vautherin in the absence of a magnetic field
\citep{Neg}. We note that our predictions for $B=4.414 \times 10^{16}$ G do
not change from the field free results because a large number of Landau levels 
is populated in that magnetic field. For magnetic field $B=10^{17}$ G,
the total number of nucleons decreases in the density regime 0.005 - 0.02 
fm$^{-3}$ compared with the corresponding results of the field free case. 
This can be understood from the behaviour of the cell size around that 
density regime in Fig. 2. For $B=4.414 \times 10^{17}$ G, the
zeroth Landau level is populated by electrons for densities $\leq 0.04$ 
fm$^{-3}$. This modifies the $\beta$-equilibrium and the charge neutrality 
conditions which, in turn, impact the size of the cell and the total number of 
nucleons in a cell. This 
effect is pronounced in the case of $B=4.414 \times 10^{17}$ G. In this case, 
$A_{cell}$ is significantly reduced compared with the zero field case 
for densities $\leq 0.04$ fm$^{-3}$.

We obtain neutron ($N$), proton ($Z$) and total nucleon numbers ($A$) in the 
nucleus at each 
average baryon density following the subtraction procedure. Total nucleon and 
proton numbers are shown in Fig. 7 for the above mentioned magnetic fields. 
When the magnetic field is $4.414 \times 10^{16}$ G or more,
It is noted that our results in certain cases start oscillating from the field 
free results.  This may be attributed to the fact that the population of Landau
levels jumps from a few levels to zero in the above mentioned fields as baryon 
density decreases from higher to lower values.  In contrast to Fig. 6, we find 
total 
nucleon and proton numbers inside the nucleus at each  density point beyond 
0.002 upto 0.04 fm$^{-3}$ are significantly enhanced in case of $B = 4.414 
\times 10^{17}$ G compared with the field 
free case as well as other magnetic fields considered here. This clearly 
demonstrates that more neutrons are 
inside the nucleus in the presence of strong magnetic fields $\geq 10^{17}$ G 
than in the gas phase in that density regime. This is opposite to the
situation in the zero magnetic field. This can be easily understood from the
density profiles with and without magnetic fields in Fig. 4 and Fig. 5.    

We plot the free energy per nucleon of the system with average baryon density 
in Fig. 8. Our results for 
$B=4.414 \times 10^{16}$ G do not change much form the field free results.
However, for $B=10^{17}$ G, the free energy per nucleon is reduced
at lower densities ($<$ 0.004 fm$^{-3}$) as compared with the field
free case. We find more pronounced reduction in the free energy per nucleon in
the field $B=4.414 \times 10^{17}$ G almost over the whole density regime 
considered here. 

\section{Summary and Conclusions}       

We have investigated properties of nuclei in the inner crust and their 
stability in the presence of strong magnetic fields $\sim 10^{16}$ or more.
Nuclei are immersed in a neutron gas and uniform background of electrons. We
have adopted the SKM$^*$ interaction for the nuclear energy density functional 
and studied this problem in the Thomas-Fermi model. 
Electrons are affected through Landau quantisation in strong magnetic fields
because much less Landau levels can be occupied in these cases. 
Consequently, electron number density and 
energy density are modified in strongly quantising magnetic field and the
$\beta$-equilibrium condition is altered compared with the field free case.
The enhancement of electron number density in magnetic fields $\geq 10^{17}$ G
due to the population of the zeroth Landau level leads to enhancement in proton
fraction through the charge neutrality condition. We minimise the free energy
of the system within a Wigner-Seitz cell to obtain the nucleus at each average
baryon density. In this connection we used the
subtraction procedure to obtain the density profiles of a nucleus from the 
nucleus plus gas and only gas solutions at each average baryon density point. 
We note that less number of neutrons drip out of a nucleus in the presence of
strong fields than the situation without magnetic field. This results in larger
mass and proton numbers in a nucleus in the presence of magnetic field 
$>10^{17}$ G compared with the corresponding nucleus in the field free case.
Further the free energy per nucleon of the system is reduced in magnetic fields 
$\geq 10^{17}$ G.

Magnetars might eject crustal matter due to tremendous magnetic stress on the
crust \citep{Gel}. The ejected matter of the inner crust might expand to much 
lower
densities. The decompressed crustal matter has long been considered as an
important site for $r$-process nuclei \citep{Schr,Ste}. It would be worth 
studying the $r$-process in the decompressed crustal matter of magnetars using
the results of our calculation as an input.

\acknowledgments

RN and DB thank the Alexander von Humboldt Foundation for the support under
the Research Group Linkage programme.

\clearpage

\begin{figure}
\epsscale{1.00}
\plotone{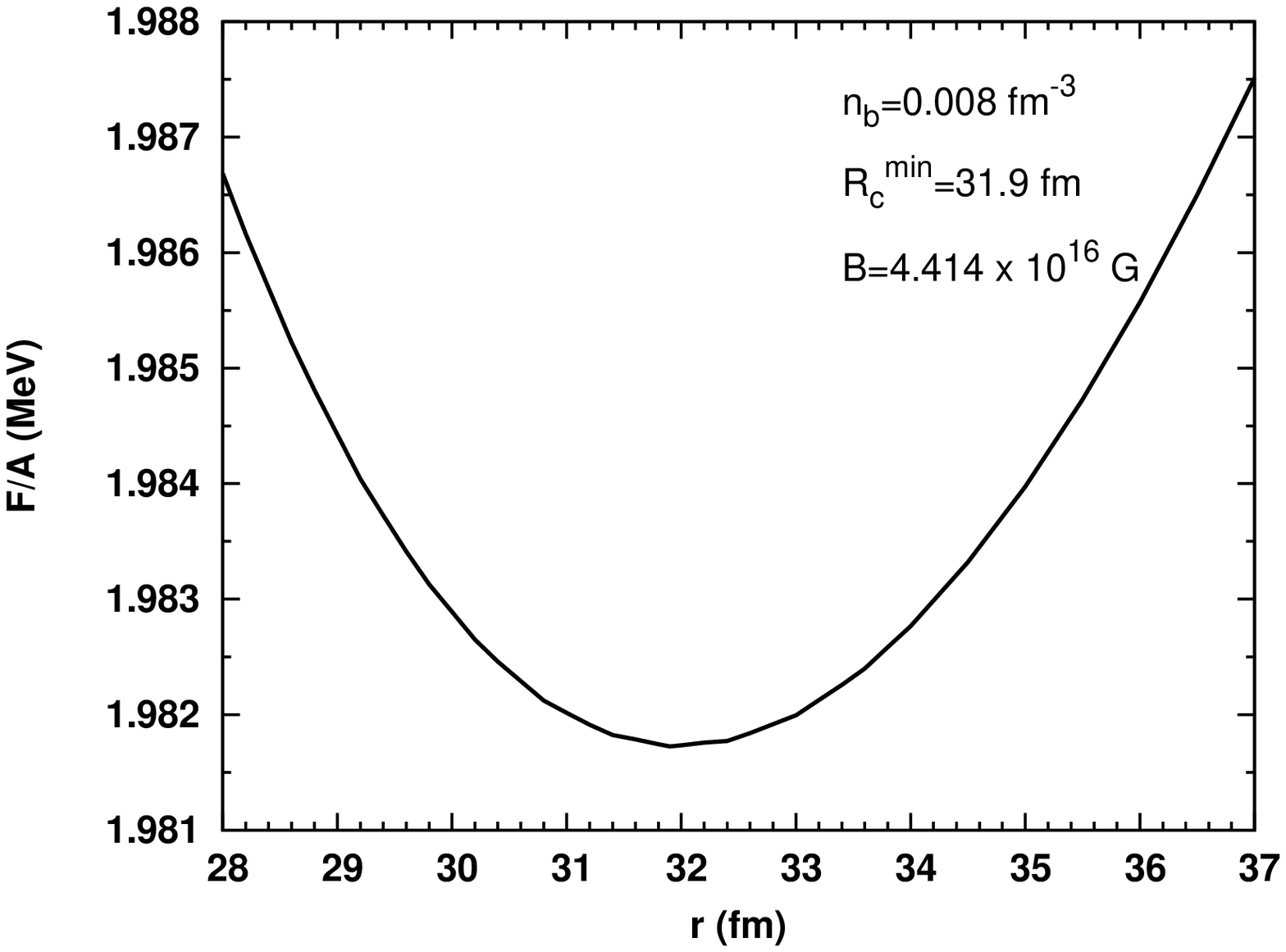}
\caption{Free energy per nucleon is plotted with cell size for average baryon 
density $n_b = 0.008$ fm$^{-3}$ and magnetic field $B = 4.414 \times 10^{16}$ 
G.}
\end{figure}

\clearpage

\begin{figure}
\epsscale{.80}
\plotone{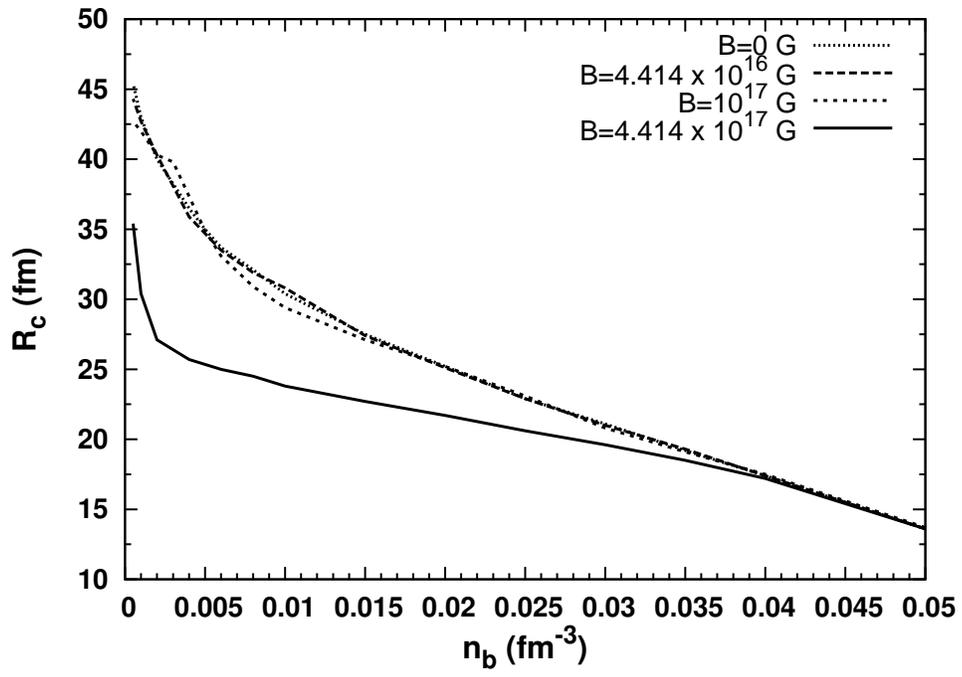}
\caption{Cell size corresponding to the free energy minimum is shown as a 
function of average baryon density for magnetic fields 
$B=0$, $4.414 \times 10^{16}$, $10^{17}$ and $4.414 \times 10^{17}$ G.} 
\end{figure}

\clearpage

\begin{figure}
\epsscale{.80}
\plotone{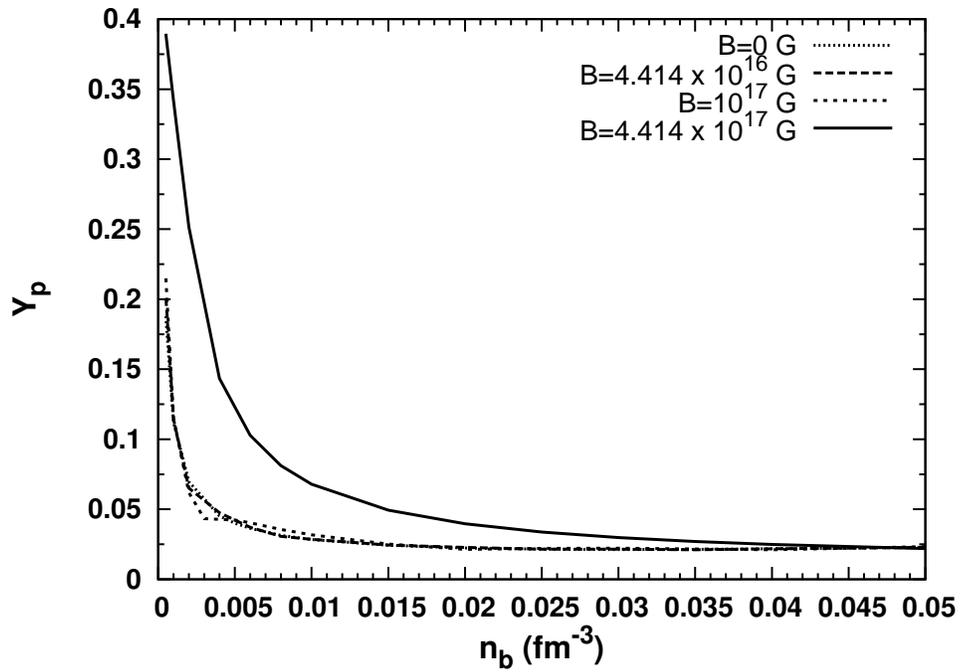}
\caption{Proton fraction is plotted with average baryon density for magnetic
fields $B=0$, $4.414 \times 10^{16}$, $10^{17}$ and $4.414 \times 10^{17}$ G.} 
\end{figure}

\clearpage

\begin{figure}
\epsscale{1.00}
\plotone{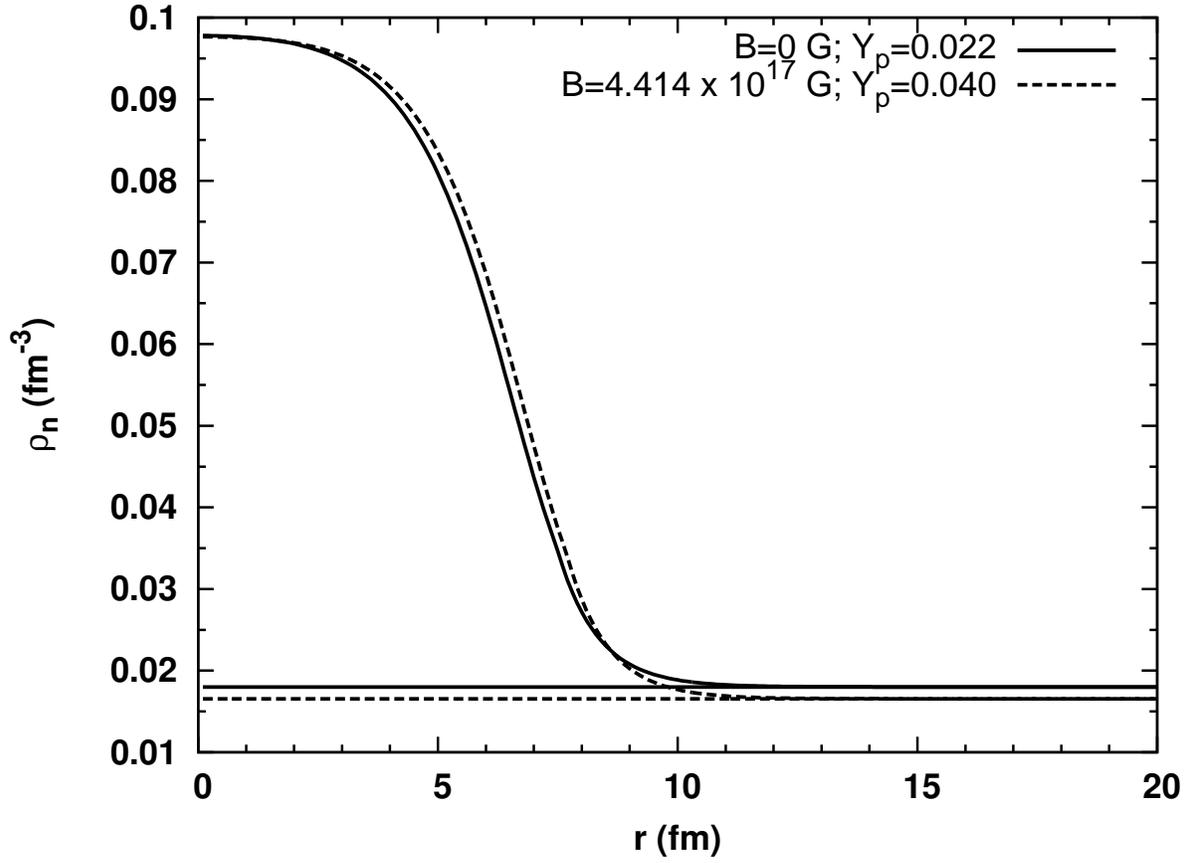}
\caption{Neutron density profile in the nucleus plus gas phase with magnetic 
field $B=4.414 \times 10^{17}$ G (dotted line) and without magnetic field 
(solid line) at an average baryon density $n_b = 0.02$ fm$^{-3}$. 
Horizontal lines denote gas phases in both cases.} 
\end{figure}

\clearpage

\begin{figure}
\epsscale{1.00}
\plotone{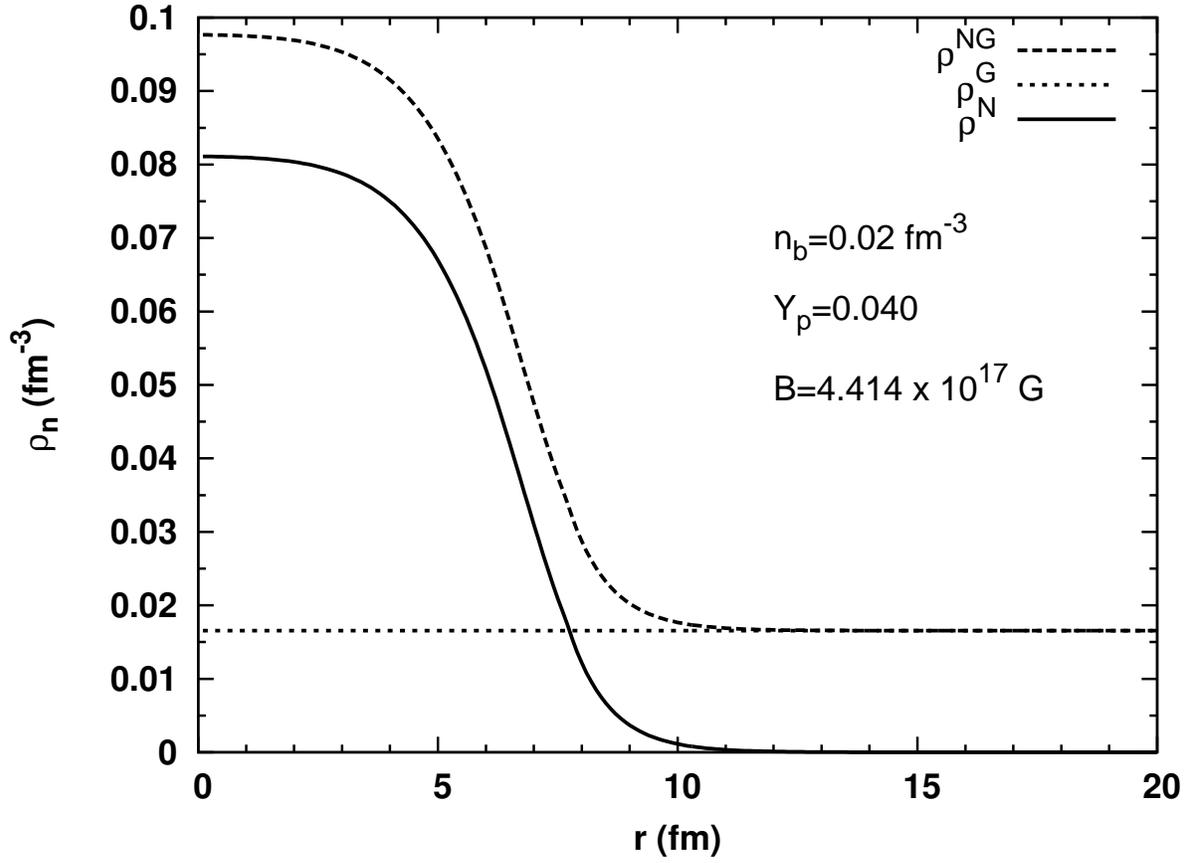}
\caption{Neutron density profile in the nucleus plus gas phase and after the
subtraction of the gas phase with magnetic field $B=4.414 \times 10^{17}$ G 
for same values of average baryon density and proton fraction as in Fig. 4.}
\end{figure}

\clearpage

\begin{figure}
\epsscale{1.00}
\plotone{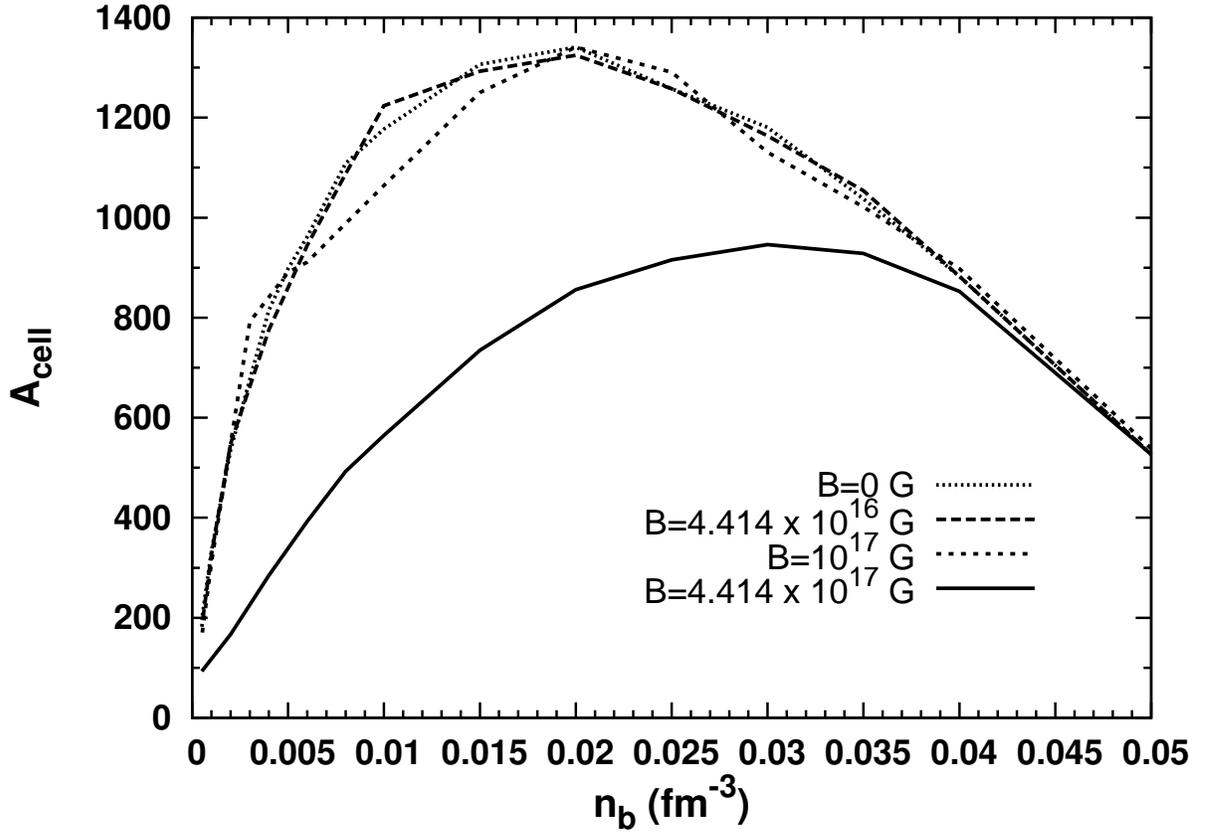}
\caption{Total number of neutrons and protons in a cell ($A_{cell}$) is 
plotted as a function of average baryon density for magnetic 
fields $B=0$, $4.414 \times 10^{16}$, $10^{17}$ and $4.414 \times 10^{17}$ G.} 
\end{figure}

\clearpage

\begin{figure}
\epsscale{1.00}
\plotone{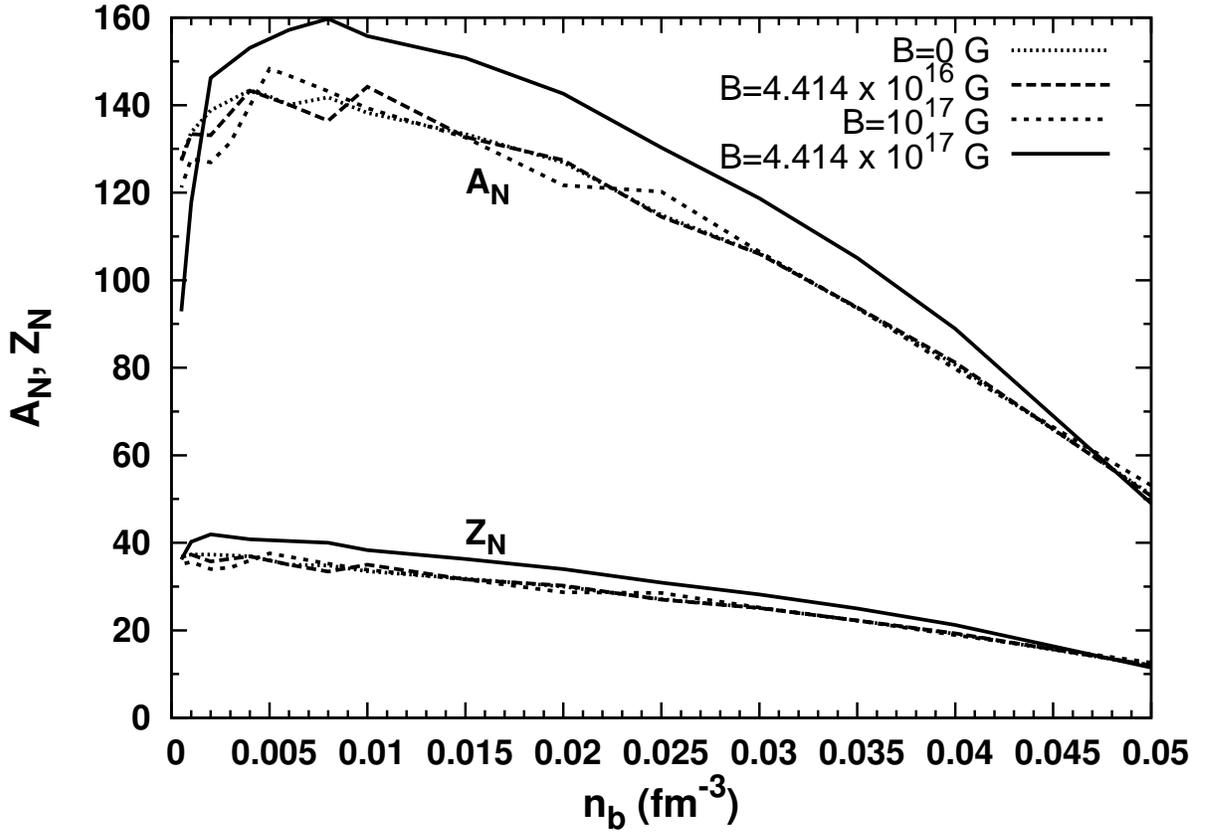}
\caption{Total nucleon and proton numbers in a nucleus are
plotted as a function of average baryon density for magnetic 
fields $B=0$, $4.414 \times 10^{16}$, $10^{17}$ and $4.414 \times 10^{17}$ G.} 
\end{figure}

\clearpage

\begin{figure}
\epsscale{1.00}
\plotone{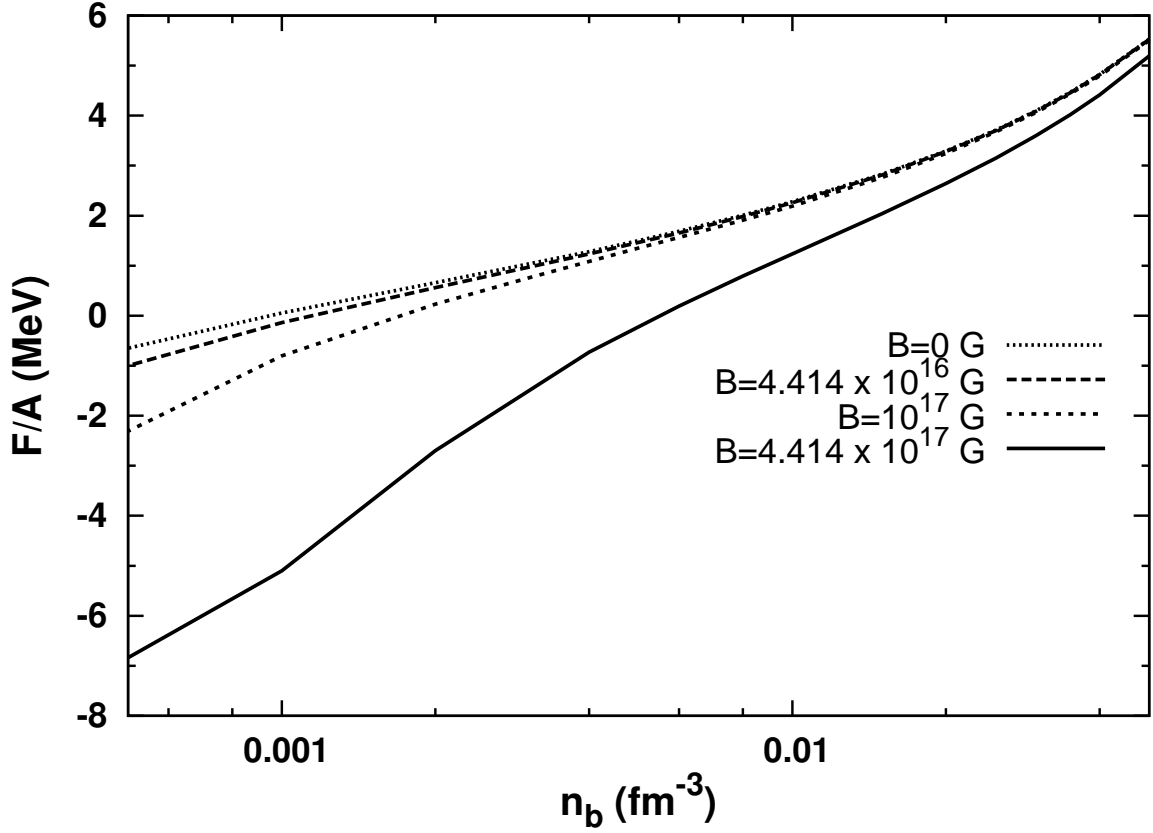}
\caption{Minimum Free energy per nucleon of the system is 
shown as a function of average baryon density for magnetic fields 
fields $B=0$, $4.414 \times 10^{16}$, $10^{17}$ and $4.414 \times 10^{17}$ G.} 
\end{figure}

\end{document}